# Entanglement, Dephasing, and Phase Recovery via Cross-Correlation Measurements of Electrons

I. Neder, M. Heiblum, D. Mahalu, and V. Umansky

*Braun Center for Submicron Research, Department of Condensed Matter Physics, Weizmann Institute of Science, Rehovot 76100, Israel*

Determination of the path taken by a quantum particle leads to a suppression of interference and to a classical behavior. We employ here a quantum 'which path' detector to perform **accurate** path determination in a two-path-electron-interferometer; leading to **full suppression** of the interference. Following the dephasing process we recover the interference by measuring the cross-correlation between the interferometer and detector currents. Under our measurement conditions every interfering electron is dephased by approximately a single electron in the detector - leading to mutual entanglement of approximately single pairs of electrons.

Experiments involving an interferometer and a path detector were already performed before in mesoscopic systems [1,2], however, path detection was not sufficiently accurate, hence leading to partial dephasing. In the present work, the detector, being populated approximately with a **single electron**, causes **total dephasing** of the interferometer. The lost interference is then being recovered with a measurement that resembles 'post selection' ('coincidence') experiments that were performed in systems with entangled photons [3,4,5]. In these experiments the usual two-slit interference was absent when a detector was scanned behind the slits, due to the photon of the *signal* beam being in spatially entangled state with the photon of the *idler* beam. The interference pattern was fully recovered only by coincidence measurements of the events in the *signal* and *idler* sides; in effect performing 'post selection' measurements. Showing and controlling such striking effects of entanglement in electronic systems is a desirable goal with the advent of quantum information.

We constructed an electronic Mach-Zehnder interferometer (MZI) [6,7], fabricated in a two dimensional electron gas (see Fig. 1), operating in the integer quantum Hall effect regime (IQHE). An edge channel is split by a quantum point contact **QPC1** to two paths, which join again in **QPC2** after enclosing a high magnetic flux. Ohmic contacts serve as sources (**S1**, **S2**, **S3**) and drains (**D1**, **D2**). Changing the flux by $\Delta\Phi$ (e.g., via biasing the modulation gate **MG)**, adds an Aharonov-Bohm (AB) phase difference $\varphi_{AB} = 2\pi \Delta\Phi / \Phi_0$ ($\Phi_0$ the flux quantum) between the paths [8]. The phase dependent transmission coefficient from source **S2** to drain **D2** is:

$$T_{MZI} = \left| t_{QPC1} t_{QPC2} + e^{i\varphi_{AB}} r_{QPC1} r_{QPC2} \right|^2 = T_0 + T_\varphi \cos\varphi_{AB}, \qquad (1)$$

where *t* and *r* are transmission and reflection amplitudes. In actual experiments the visibility is $\nu = T_\varphi / T_0 = 30-60\%$ [6,7]; lower than unity, most likely, due to fluctuations in the enclosed flux that may result from external noise sources [9]. We return to this point later. Note that a change in area of only some 100nm$^2$ suffices to change the AB phase by $2\pi$.

The detector was constructed as follows. Tuning the magnetic field to filling factor 2 in the IQHE, two edge channels, an outer and an inner, were injected from sources **S2** and **S3**. **QPC0** was tuned to fully transmit the outer channel and partly reflect the inner one. Consequently, two channels impinged on **QPC1** from the right: a fully occupied outer channel from **S2** and a partitioned inner channel from both **S2** and **S3**. In turn, we tuned $T_{QPC1}=T_{QPC2}=0.5$ (0) for the outer (inner) channel. Consequently, while the outer channel interfered, the inner one flowed in close proximity to the upper path; serving as a 'which path' detector.

Dephasing, or path detection, results from Coulomb interactions between electrons in the inner and outer channels. Electrons in the two channels repel each other and affect each other's phase. The interaction between the inner and outer channels can be quantified experimentally by fully reflecting the inner channel, with **QPC0**, so that it arrives as a full and noiseless beam from **S3**, serving as a biased 'gate' to the upper path, with $V_{det}=V_{S3}-V_{S2}$. As $V_{det}$ was changed (by fixing $V_{S2}\sim 0$ and changing $V_{S3}$), so was the enclosed area between the two paths, resulting in a linear phase shift of the AB oscillations of $2\pi$ per $\Delta V_{det} \approx 20\mu V$ (see Fig 2 and Ref. 10). However the detector bias didn't influence the QPCs transmission amplitudes $t_{QPC1}$ and $t_{QPC2}$ (not shown), hence it had an insignificant effect on the amplitude of the oscillations (Fig 2a).

Partitioning the inner channel (via un-pinching **QPC0**), splits the detector beam into two paths, providing a reference path for an 'in principle' interference experiments of the detector states. Such experiment, if done, would measure the phase the detector acquires due to the presence of an electron in the upper path of the MZI; hence, detection. The same phase, by symmetry, is the phase induced in the interferometer. Formally, after the interaction, the interferometer & detector wave function is an entangled state:

$$|\Psi\rangle = |\psi_l\rangle \otimes |\chi_l\rangle + e^{i\varphi_{AB}} |\psi_u\rangle \otimes |\chi_u\rangle , \qquad (2)$$

where $|\psi_{l,u}\rangle$ are the interferometer's partial wave functions with an electron in the lower or upper paths, $|\chi_{l,u}\rangle$ are the corresponding detector wave functions, and $\varphi$ is the AB phase. If the outgoing wave function of an electron in drain **D2** is $|D_2\rangle$, the probability to find the interfering electron in **D2** is $P = |\langle D_2|\Psi\rangle|^2$, namely:

$$P = |\langle \psi_l|D_2\rangle|^2 + |\langle \psi_u|D_2\rangle|^2 + 2\,\text{Re}\,[e^{i\varphi_{AB}} \langle \psi_l|D_2\rangle\langle D_2|\psi_u\rangle \cdot \langle \chi_l|\chi_u\rangle]. \qquad (3)$$

The overlap of the detector states that multiply the interference term determines the visibility. For an absolute path determination, and a vanishing interference, the two detector states must be orthogonal.

The two detector's states $|\chi_{l,u}\rangle$ are generally complicated many-electron-states [7], however, here we consider a simple model with *n* independent electrons in the detector channel interacting with an electron in the interferometer. Each of the detecting electron states can be expressed as follows: $|\chi_l\rangle = t_d|t_d\rangle + r_d|r_d\rangle$ and $|\chi_u\rangle = t_d|t_d\rangle + e^{i\gamma} r_d|r_d\rangle$, where $t_d|t_d\rangle$ and $r_d|r_d\rangle$ are the amplitudes of the

transmitted and reflected electron wave functions (by **QPC0**), and $\gamma$ the induced phase in the detector in the presence of an electron in the upper path of the interferometer. The visibility then is [1,2,11]:

$$\left|\langle \chi_l | \chi_u \rangle\right|^n = \left(1 - 4t_d^2 r_d^2 \sin^2 \frac{\gamma}{2}\right)^{\frac{n}{2}}, \qquad (4)$$

with $n \sim \frac{I_{det}\tau_d}{e} = \frac{eV_{det}\tau_d}{h}$, $\tau_d$ dwell time of an electron in the upper path of the MZI, and $\frac{e^2}{h}$ the channel conductance. As $\gamma$ or $V_{det}$ increase, a more accurate path detection can be achieved. For a phase shift $2\pi$, at $V_{det}=20\mu V$, we find $n\sim 1$-$2.5$ (using $\tau_d = \frac{L}{v_g}$, path length $L\approx 10\mu m$, and $v_g \sim (2-5)\cdot 10^6$ cm/sec as rough estimate of the edge channel drift velocity). Note that such strong interaction is in contrast with previous experiments [1,2], where $\gamma \ll \pi$ and dephasing resulted from many 'weakly detecting electrons'.

In Fig. 3 we show the effect of a partitioned detector (inner channel, $T_{QPC0}\sim 0.5$) on the interference of the MZI. For poor path detection ($V_{det}=2\mu V$) the AB oscillations are strong with visibility $\sim 30\%$ (Fig. 3a), but for an accurate detection ($V_{det}=24\mu V$) the visibility drops to merely $\sim 1.5\%$ (Fig. 3b); vanishing altogether as $V_{det}$ increases further (Fig. 3c). Moreover, the dependence of the visibility on the partitioning $T_{QPC0}$ (Fig. 3c, upper inset) changes from a parabolic like to a sharp V-shape like as $V_{det}$ increases to $15\mu V$, with a minimum near $T_{QPC0}=0.5$. It is interesting to note that for $n=1$ and $\gamma \sim \pi$, Eq. 4 indeed leads to a V-shape; $\nu \propto |1-2T_{QPC0}|$. This is a direct consequence of higher moments of the shot noise in the detector edge channel.

Measuring the second moment (lower inset in Fig 3c, $T_{QPC0}$=0.5) obeyed the predicted noise $S_{det}=2eV_{det}(e^2/h)T_{QPC0}(1-T_{QPC0})$ with finite temperature corrections [12], proving indeed that stochastically partitioned electrons produced the shot noise.

Can the lost interference be recovered? Dephasing the MZI results from averaging over the detector states, namely, over the presence of an electron (probability $R_{QPC0}$) or its absence (probability $T_{QPC0}$). However, if we were to select, among the interfering electrons, only these passing simultaneously with a detector electron, these ones would acquire the same phase $\varphi_{AB}+\gamma$ and interfere. We show now that a cross-correlation measurement between current fluctuations in the detector and in the MZI 'posts selects' the interfering electrons in our configuration. For this purpose we measure the total current fluctuations in **D2**:

$$\langle(\Delta n_{D2})^2\rangle=\langle(n_{MZI}+n_{det})^2\rangle-(\langle n_{MZI}\rangle+\langle n_{det}\rangle)^2=\langle(\Delta n_{MZI})^2\rangle+\langle(\Delta n_{det})^2\rangle+2\langle\Delta n_{MZI}\cdot\Delta n_{det}\rangle, \quad (5)$$

containing, the uncorrelated sum of the MZI noise, the detector noise, and the sought after cross-correlation (CC) term. In a fully dephased MZI only the CC term is expected to be phase dependent.

We start first with an unbiased detector and slightly biased MZI. For $V_{det}$=0 ($\Delta n_{det}$=0), the measured zero temperature total noise in **D2** is that of the interferometer $S_{D2}= S_{MZI} = 2eI_{imp}T_{MZI}(1-T_{MZI})$, with $I_{imp} = \frac{e^2}{h}V_{S2}$. In a symmetric interferometer $T_0$=0.5, $T_\varphi$=0.5$\nu$ and $S_{MZI} = 0.5eI_{imp}(1-\nu^2\cos^2\varphi_{AB})$; namely, shot noise with only a **second AB harmonic**. Biasing the MZI with $V_{S2}$=4.5μV (with a negligible detector bias), we show in Fig. 4a short segment of the oscillating total noise. Compared with the current oscillations in Fig. 3a, the noise oscillated at half the AB period - shown by the Fourier transform in the inset. Note, however, that the amplitude of the oscillations in the noise was five times stronger than the expected one (with $\nu$~45% at $V_{S2}$=0). We account this discrepancy to an external (unavoidable) noise, that leads to fluctuations in the AB phase, and hence, in the transmission $\Delta T_{MZI} \propto \Delta(T_\varphi \cos\varphi_{AB})$. The resultant current fluctuations in the MZI, being proportional to $(\Delta T_{MZI})^2$, oscillate only with the **second AB harmonic**. Hence, this noise measurement provides new clue regarding the lower visibility (45% instead of 100%). The MZI is most likely subjected to some low-frequency noise that causes 'classical phase averaging', while each passing electron stays coherent.

Before performing the CC measurement, we argue that for large enough $V_{det}$ that causes total dephasing, the CC term in a simple model of **one** electron in the detector coupled to **one** electron in the MZI, recovers the 'lost interference'. In the fully dephased MZI, the oscillating part of the CC term $\langle \Delta n_{MZI} \cdot \Delta n_{det} \rangle = \langle n_{MZI} \cdot n_{det} \rangle - \langle n_{MZI} \rangle \cdot \langle n_{det} \rangle$ is given by: $\langle n_{MZI} \cdot n_{det} \rangle = \sum_{n_{MZI}, n_{det}=0,1} P(n_{MZI}|n_{det}) P(n_{det}) n_{MZI} n_{det}$, with $P(n_{MZI}|n_{det})$ the conditional probability for $n_{MZI}$ (0 or 1) interfering electrons arriving at **D2** provided that $n_{det}$ detecting electrons arrive simultaneously at **D2**. The only non-vanishing term is $n_{MZI} = n_{det} = 1$, with $P(n_{det}=1)=R_{QPC0}=0.5$ and $P(n_{MZI}=1|n_{det}=1)=T_0+T_\varphi \cos(\varphi_{AB}+\gamma)$. Hence, for $n$=1, $\gamma=\pi$, $T_0$=0.5 and a pre-dephased visibility $\nu$, the spectral density of the CC term is $S_{CC}=A\,\nu\cos(\varphi_{AB}+\gamma)$, with $A$ prefactor. In other words, the CC term is the only term in the total noise with the **basic AB periodicity**.

The prefactor *A* can be roughly estimated in our "*n*=1" model. One can simplify it further by assuming that the shot-noise in the detector behaves as switching noise, in the sense that the potential in the detector fluctuates randomly every time interval $\frac{e}{I_{det}} = \frac{h}{eV_{det}}$ between two possible values: $V_{det}$ when an electron is present with probability $R_{QPC0}$, and 0 when it is absent with probability $T_{QPC0}$. As a result the current at the output of the MZI fluctuates between $V_{MZI} = V_{S2}T_0[1 + v\cos(\varphi_{AB} + \gamma)]$ when the detector bias is $V_{det}$, and $V_{MZI} = V_{S2}T_0[1 + v\cos(\varphi_{AB})]$ when the detector bias is 0. A standard calculation shows that the CC term in this case is:

$$S_{CC} = 2\langle \Delta i_{MZI} \cdot \Delta i_{det} \rangle_{F.T., f \to 0} = 8eI_{S2}T_0 v R_{QPC0} T_{QPC0} \sin(\gamma/2)\cos(\varphi_{AB} + \gamma/2). \quad (6)$$

At low detector bias Eq. (6) reduces to the first order effect of the detector's current fluctuations on the current of the MZI via a classical like *transconductance* term,

$$S_{TC} = 2\langle \Delta i_{MZI} \cdot \Delta i_{det} \rangle = 2\left\langle \frac{di_{MZI}}{di_{det}} \Delta i_{det} \cdot \Delta i_{det} \right\rangle = 2\frac{di_{MZI}}{di_{det}} \langle \Delta i_{det}^2 \rangle,$$

with the detector shot noise $\langle \Delta i_{det}^2 \rangle$ known, linearly dependent on $V_{det}$ (Fig. 3c, lower inset), and $\frac{di_{MZ}}{di_{det}}$ is proportional to the *transconductance* through the phase dependence on the detector voltage, $\frac{di_{MZI}}{di_{det}} = \frac{dV_{MZI}}{dV_{det}} = \frac{dV_{MZI}}{d\varphi} \cdot \frac{d\varphi}{dV_{det}} = V_{S2}T_0 v \cdot \sin\varphi_{AB} \cdot \frac{2\pi}{20\mu V}$. In other words, for $T_0 = 0.5$, $R_{QPC0} = T_{QPC0} = 0.5$, the cross-correlation should rise linearly with $V_{det}$, reaching ~1·10$^{-29}$ A$^2$/Hz at $V_{det}$=5μV.

At higher detector bias, e.g., when $\gamma=\pi$ (when the interference in the conductance vanishes), the first order estimate is not valid any more. Equation (6) predicts that the AB oscillations in the CC term **reach a maximum** $S_{CC} \sim 1.3 \cdot 10^{-29} A^2/Hz$; somewhat smaller than our measurement. Note that $S_{CC}$ depends on the **undephased** visibility ($v$=45%) – proving that original phase dependence is indeed recovered. This can be better understood from our above argument of $\langle \Delta n_{MZI} \Delta n_{det} \rangle$, namely, we perform 'post selection' of only the interfering electrons that sensed the presence of an electron in the detector channel, hence, all these selected electrons have the same phase.

The evolution of the noise with increasing $V_{det}$ was then measured. A sample of such data is shown in Fig. 4b for an almost fully dephased interferometer ($V_{S2}$=4.5µV, $V_{det}$=24µV), with the total noise in **D2** oscillating now **only** at the basic AB frequency. A summary of the evolution of the two AB harmonics in the noise as function of $V_{det}$ is given in Fig. 5. As $V_{det}$ increased the second AB harmonic decreased too and vanished altogether – a consequence of the quenched phase dependent transmission of the MZI. The first AB harmonic, though, appeared fast, saturated, and then grew abruptly with $V_{det}$. After reaching a peak it fell and vanished near $V_{det} \cong 34$µV. The strength of the CC term agrees roughly with a simple estimate above (Note that the estimate is rather rough, since it doesn't consider the exact dynamics of the detecting and interfering electrons, such as their velocities, the finite range of Coulomb interaction, and the entering of other electrons into the detector). The reasons for the sudden increase in the CC term ($V_{det}$~15-18µV) and the subsequent decrease ($V_{det}$>25µV) are not clear. The deviation from the simple estimate of the CC term may be related to an onset of a second electron occupying the detector channel. The decrease of the second AB harmonic at high $V_{det}$ could result from new degrees of freedom opening up in the detector - unaccounted for in our model.

Here we exploited the strong interaction between electrons in adjacent edge channels to strongly entangle approximately single pairs of electrons - each electron in a separate channel [14,15]. While the interference vanished, via measuring the cross-correlation between the current fluctuations in the two channels, detector and interferometer, phase information was recovered in some range of parameters. A more direct proof of entanglement [16] (i.e., testing Bell inequalities) can be provided if the detector edge channel were also made to interfere, say, in another MZI.

**Acknowledgement**

We are indebt to N. Ofek and A. Ra'anan for their valuable help in the experiment; to Y. Gefen, Y. Levinson, A. Stern and F. Marquardt for helpful discussions. The work was partly supported by the Israeli Science Foundation (ISF), the Minerva foundation, the German Israeli Foundation (GIF), the German Israeli Project cooperation (DIP) and the Ministry of Science - Korea Program.

**Figure Captions**

**Figure 1.** The scanning electron microscope micrograph of the actual MZI and the detector. The inner contact (**D1**) and the two QPCs are connected via air bridges. The edges of the sample are defined by etching of the GaAs-AlGaAs heterostructure, embedding a high mobility 2D electron gas. An applied perpendicular magnetic field of ~3T leads to filling factor 2 in the bulk (electron temperature ~15mK). The signal at **D2** is filtered with a cold LC resonant circuit, with center frequency ~1MHz and bandwidth ~30kHz. It is amplified in situ by a low noise preamplifier cooled to 4.2K.

**Figure 2**. The effect of an unpartitioned inner channel ($R_{QPC0}=1$) on the amplitude (Fig.2a) and phase (Fig.2b) of the outer, interfering, channel. The inner channel, injected from **S3**, is fully reflected by **QPC0**. It flows then parallel to the interfering upper path and shifts its phase. The interfering channel is weakly DC biased (VS2=4.5µV) with an additional AC signal ~1µV, ~1MHz. The amplitude of the AC signal, monitored at D2, oscillates as function of the voltage on the modulation gate (via the AB effect) with visibility ~35%. The phase is highly sensitive to $V_{det}$, $\frac{d\varphi}{dV_{det}} \cong \frac{2\pi}{20\mu V}$, while the visibility of the interference oscillations remains constant with $V_{det}$.

**Figure 3.** Path determination leading to dephasing. **a.** With negligible voltage applied on the detector channel ($V_{det}=V_{S3}-V_{S2}\sim 2\mu V$), $T_{MZI}$ oscillates as function of the modulation gate voltage with differential visibility of ~35% (here $V_{S2}=4.5\mu V$, while at $V_{S2}=0$ the visibility reached 45%). The Fourier transform has a peak at ~0.8 oscillation per 1mV (inset). **b.** At $V_{det}=24\mu V$ the interference oscillations drop by more then an order of magnitude. **c.** The evolution of the visibility as function of $V_{det}$. The small increase of the visibility near $V_{det}=5\mu V$ is related to resonances in the conductance of the **QPC0** at small $V_{det}$. **Upper Inset:** The dependence of the visibility on $T_{QPC0}$ for $V_{det}=6\mu V$ (gray dots) and for $15\mu V$ (black dots). The scattering of the data are a result of the resonances in **QPC0**. The doted line is the theoretical result of Eq. 4 with $\gamma=\pi$ and $n=1$. **Lower Inset:** The dependence of the noise in the detector on $V_{det}$ compared with the prediction of the shot noise due to independent electrons at T=15mK.

**Figure 4.** Oscillatory component of the total noise. Applying a DC bias on **S2** $V_{S2}=4.5\mu V$ and varying $V_{S3}$, the total noise is measured at **D2** (~1MHz, ~30kHz bandwidth). It contains the contributions of the MZI, the detector, and their cross correlation. **a.** For $V_{det}\sim 2\ \mu V$ there is no dephasing. The noise is mostly that of the MZI (plotted around the average), having only a second AB harmonic. **b.** With $V_{det}=24\mu V$, the AB oscillations of the conductance nearly vanish (Fig. 3b), however, the shot noise oscillates at the basic AB periodicity (plotted around the average).

**Figure 5.** Evolution of the AB oscillations in the total noise as function of the detector voltage $V_{det}$. **a.** The strength of the first AB harmonic. **b.** The strength of the second AB harmonic. As the second harmonic vanishes with the dephasing process, the basic AB harmonic rises slowly, then rapidly to a maximum, and subsequently falling and disappearing around detector bias $V_{det}\sim 34\mu V$. Both the sharp peak and the vanishing of the basic AB harmonic are not understood (see text).

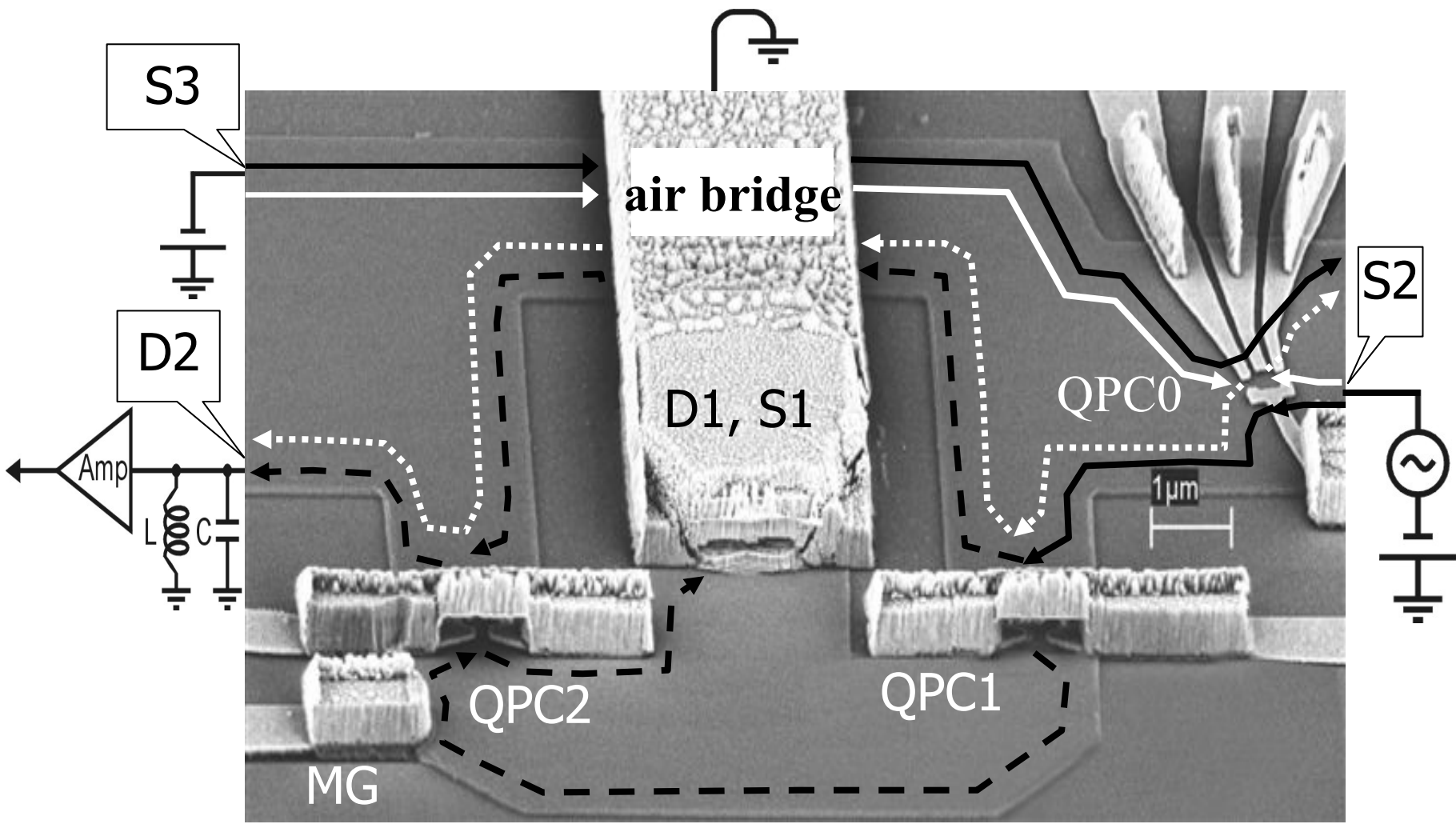

Fig. 1

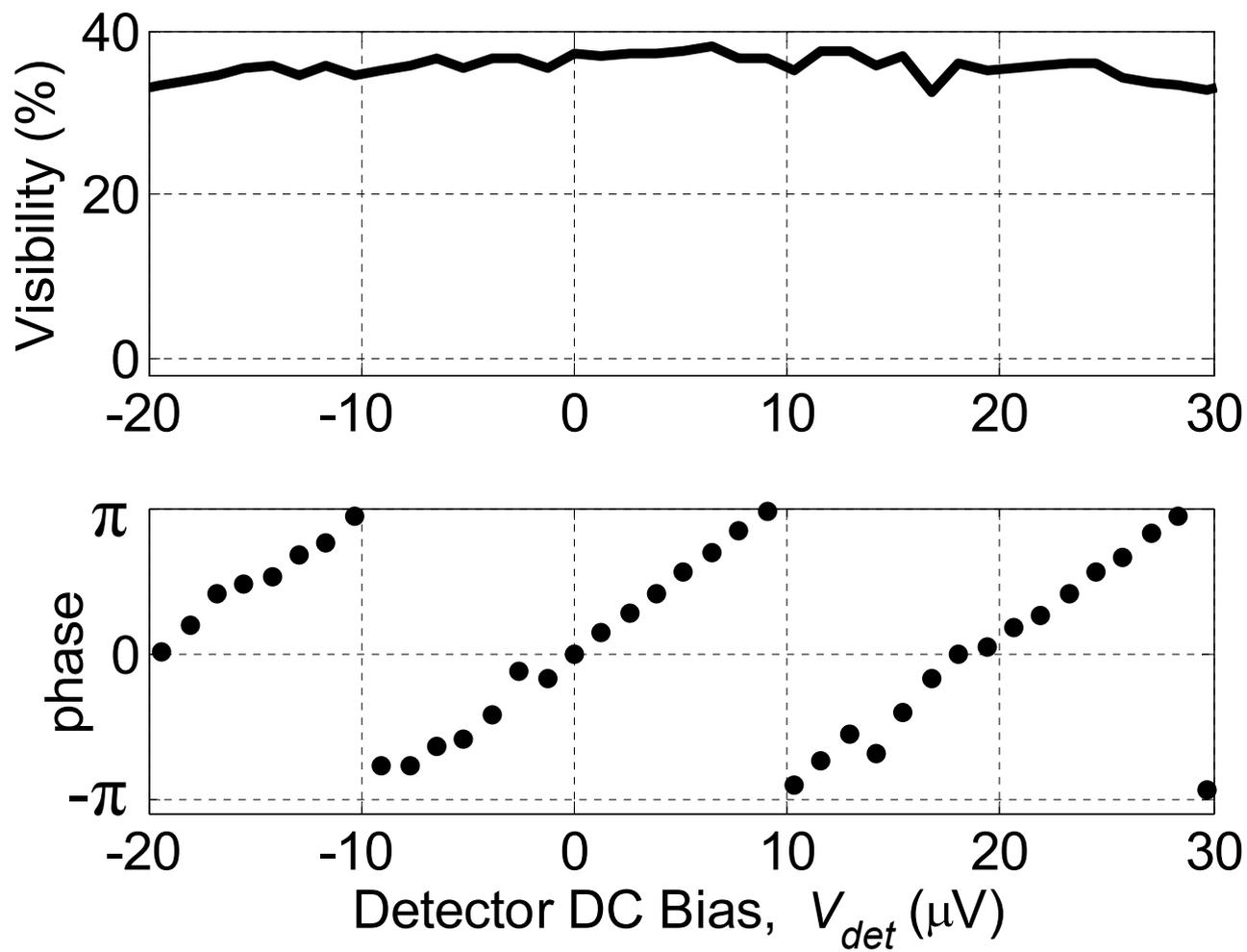

Fig. 2

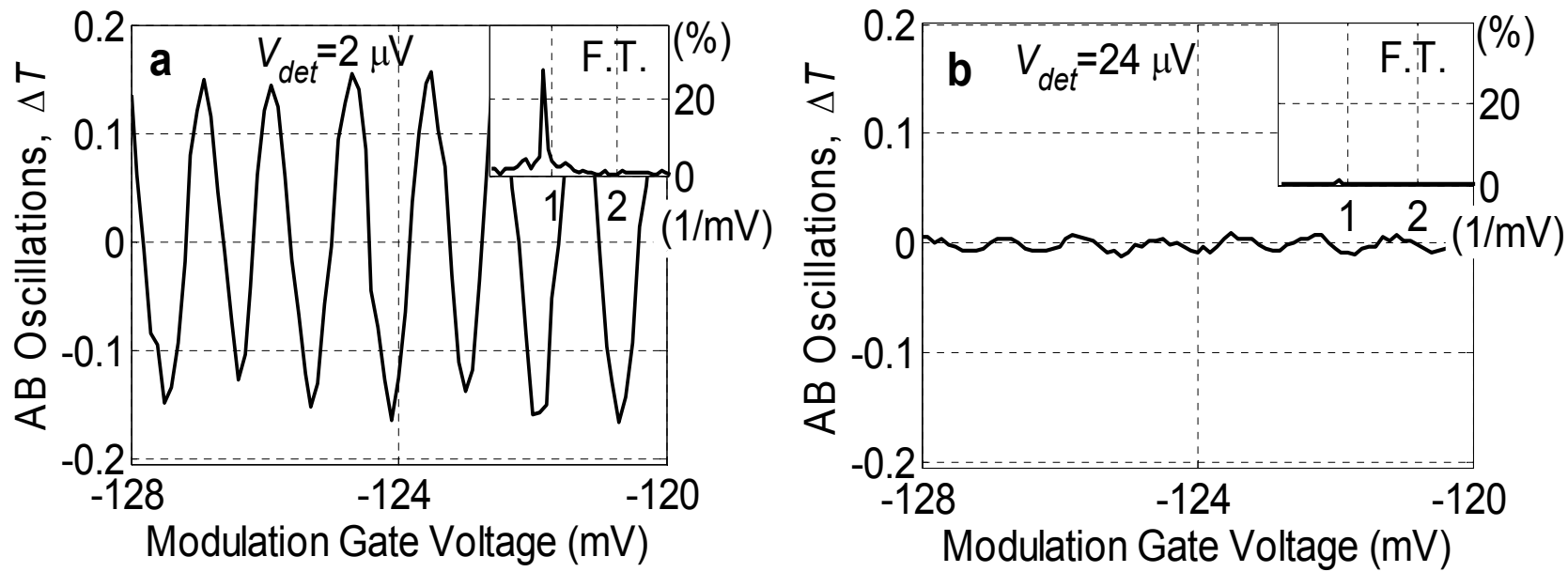
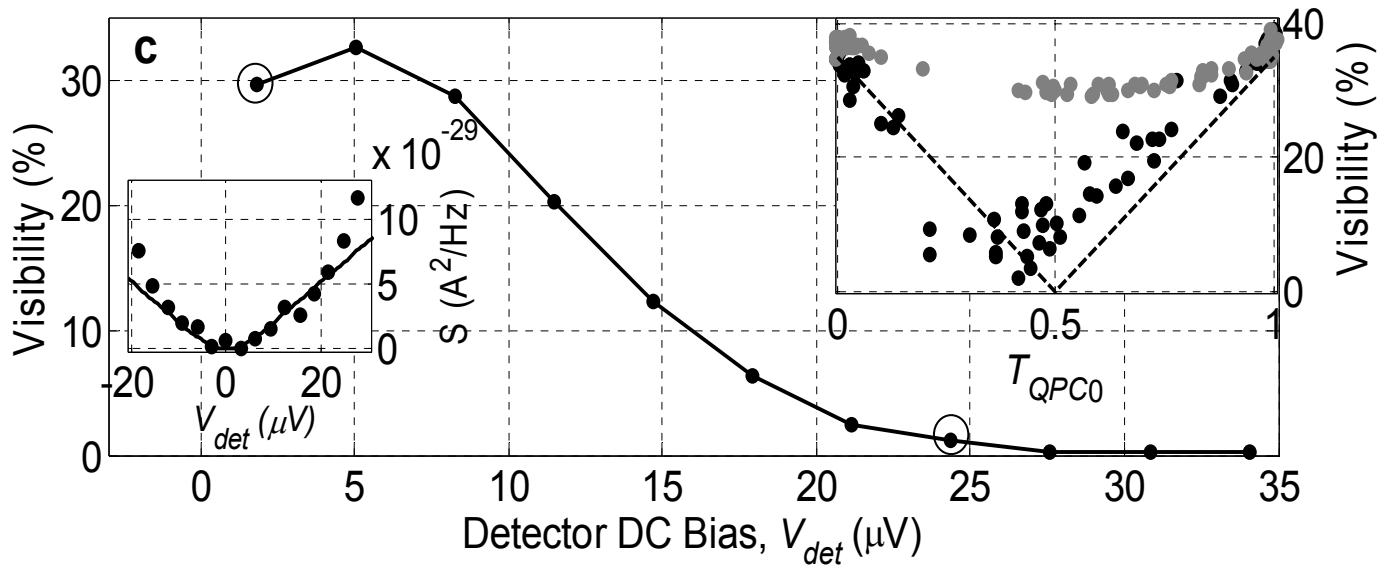

Fig. 3

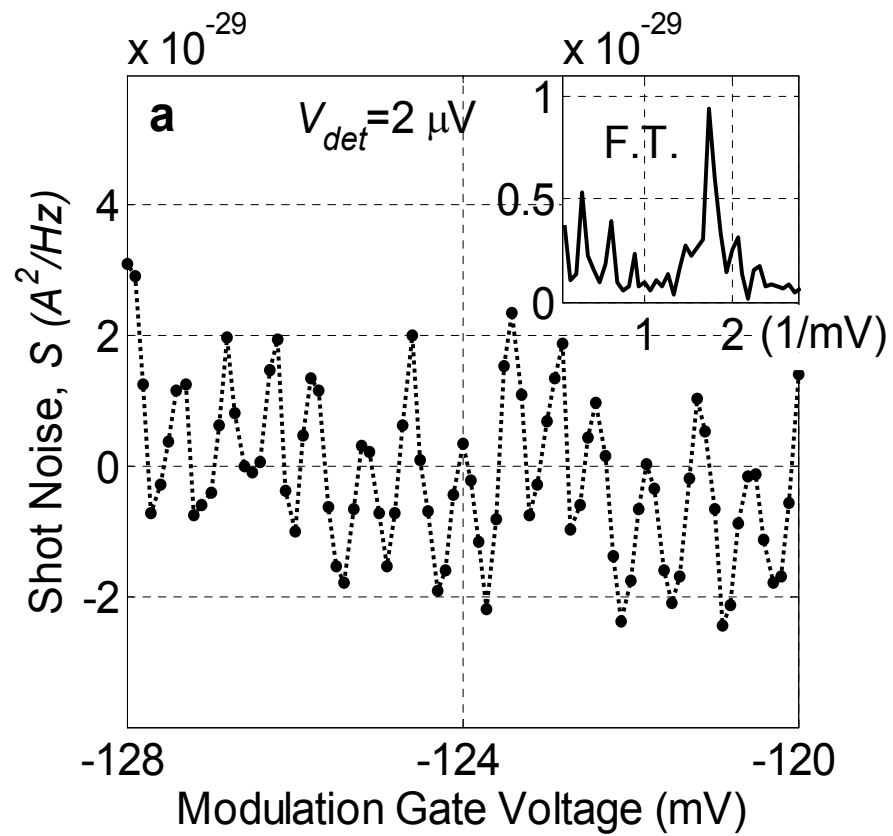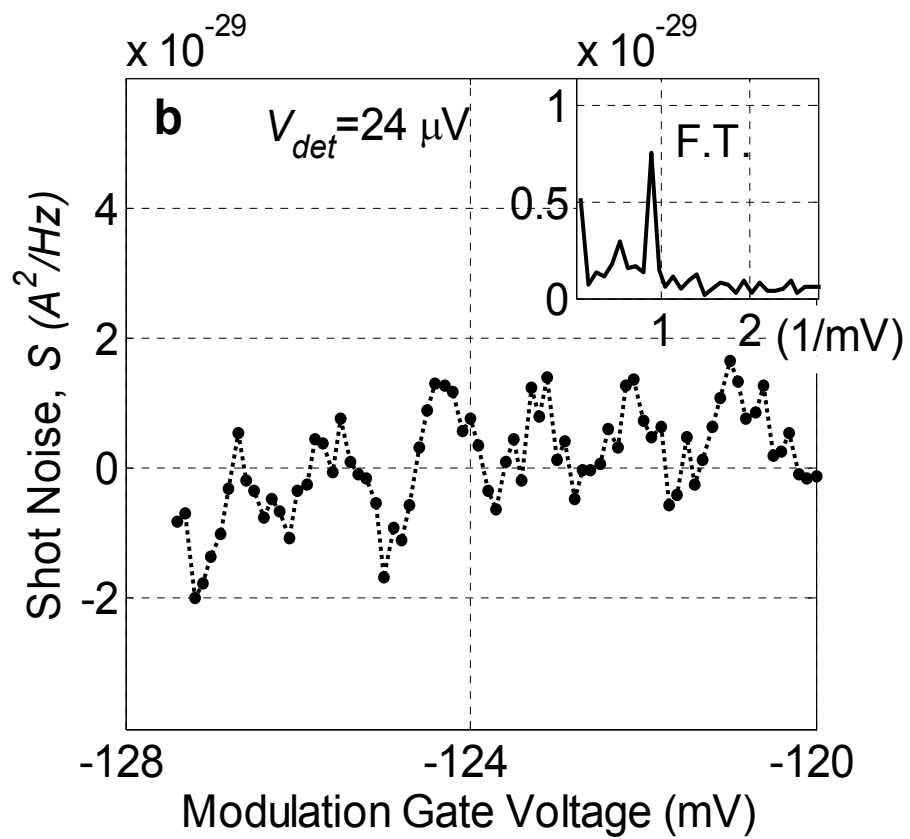

Fig. 4

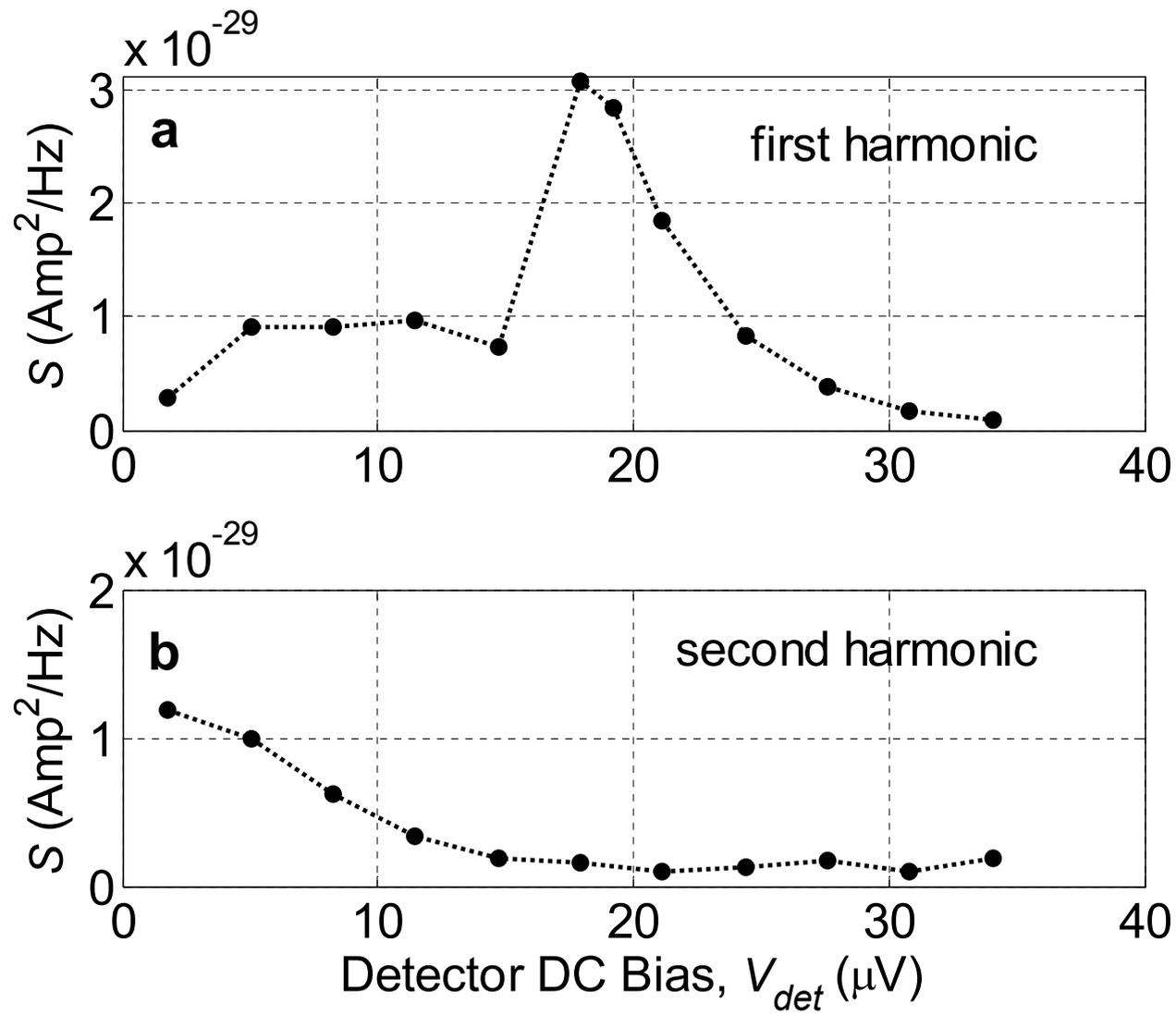

Fig. 5